\documentclass{article}
\usepackage{graphicx} 

\usepackage{amsmath}
\usepackage{amsfonts}
\usepackage{amssymb}
\usepackage{amsthm} 
\usepackage{color}
\usepackage{enumerate}
\usepackage{epstopdf}
\usepackage{multirow}
\usepackage[T1]{fontenc}
\usepackage{framed}
\usepackage[breakable]{tcolorbox}
\usepackage{hyperref}
\usepackage[latin1]{inputenc}
\usepackage{geometry}
\usepackage{paralist} 

\numberwithin{equation}{section}
\theoremstyle{definition}
\newtheorem{ex}[equation]{Example}	
\newtheorem{attack}{Attack}

\title{Chunking Attacks on File Backup \\ Services using Content-Defined Chunking}
\author{Boris Alexeev \\\\ Colin Percival \\ Tarsnap Backup Inc. \\\\ Yan X Zhang \\ San Jos\'e State University}

\date{March 2025}

\begin{document}
\maketitle
\begin{abstract}
    Systems such as file backup services often use \emph{content-defined chunking} (CDC) algorithms, especially those based on \emph{rolling hash} techniques, to split files into chunks in a way that allows for data deduplication. These chunking algorithms often depend on per-user parameters in an attempt to avoid leaking information about the data being stored. We present attacks to extract these chunking parameters and discuss protocol-agnostic attacks and loss of security once the parameters are breached (including when these parameters are not setup at all, which is often available as an option). Our parameter-extraction attacks themselves are protocol-specific but their ideas are generalizable to many potential CDC schemes.
\end{abstract}

\section{Introduction}

Online file backup services are one of several related applications that involving storing and manipulating large amounts of changing data.  One useful feature of such a service is \emph{deduplication}: if the same file is uploaded more than once, then perhaps storing it the second time should be relatively cheap.\cite{Xia2016Comprehensive}  Such a system is even more useful if uploading a slightly-modified file also resulted in cost savings, and luckily this is possible using the clever idea of \textbf{content-defined chunking} (CDC).

Some file backup services advertise the further feature that the backup service itself cannot read the user's data.  This causes some tension with respect to the deduplication feature: how can the service deduplicate data without being able to see it?  There is not yet a universally-used solution to this problem, so each backup service tends to implement their own method of performing content-defined chunking using user-specific secret keys.  In this paper, we discuss some methods of attacking these schemes in order to extract the secret keys and then use this information to recover some of the user's data.  There are also interesting interactions with another common feature of backup services: \emph{compression}.

\begin{samepage}
We begin by defining a \emph{chunking} scheme as an algorithm where:
\begin{itemize}
\item The input is a string $S$ of \emph{characters} taking the form $b_1 b_2 b_3 \cdots$, where each character $b_i$ is an element of some \emph{alphabet} set $B$. For example, if the $b_i$'s are bytes (as is typical for our use cases), then we can consider $B = \{0, \ldots,255\}.$
\item The output is a list of \emph{chunks} $C_1, C_2, \ldots$, which are themselves lists of characters, such that $S = C_1||C_2||C_3||\cdots$ where $||$ denotes concatenation. We say that $S$ \emph{chunked at} position $j$ (or equivalently, $j$ is a \emph{breakpoint}) if $b_j$ is the last character in some $C_i$. 
\end{itemize}
\end{samepage}

Content-defined chunking schemes are chunking schemes where the breakpoints are determined by the content of the surrounding data (typically the data immediately preceding the breakpoint). An archetypal application of CDC is for a data backup service: if a server is storing data for a user, then CDC enables a slight update of the data to only affect a small number of chunks near the edit, whereas a more naive chunking scheme might not. For example, if the chunking scheme simply divided the data into fixed chunk sizes, then any edit that changes the length of the document at a single place (like a one-character insertion) would require all chunks after the edit point to be remade, which might be a huge overhead requiring resending many chunks to the server.\cite{Williams1999Partitioning}

We mostly restrict our attentions to CDCs that follow a \emph{rolling hash} format (or something very similar, such as the scheme used by Tarsnap), meaning:
\begin{itemize}
\item As (public or private) parameters, we have some ring $R$, such as $\mathbb{Z}/n\mathbb{Z}$ or $F_2[X]$ and some \emph{window size} $N$. Typically we want $N$ to be much smaller than the average chunk size, so that two strings with small edit distance will end up being divided into mostly-identical sequences of chunks.
\item As we read in characters, we chunk after a string ending with $b_1 \cdots b_N$ if $$\sum_{i=1}^N g(i) f(b_i) \in R_C,$$ for some functions $g$ and $f$ (usually $g$ is some function that is easy to iterate, such as taking a power or multiplying by an element of some cyclic group) and some fixed subset of elements $R_C \subset R$. When this happens, we say that there was a \emph{clash} after $b_N$. 
\item We also chunk if the current ``unchunked'' part of the buffer since the last breakpoint (or the beginning of the file) has reached some maximum chunk size parameter (and/or the end of the data to be chunked). Thus, we can think of a clash as a special case of a chunk being created that corresponds to some algebraic constraint in $R$. This allows us to assume that we know (with high probability) if a chunk happened due to a clash or not by just looking at the context.
\end{itemize}
Some popular examples of rolling hashes frequently used in practice are the Rabin-Karp algorithm's rolling hash \cite{rabin-karp} (modulo some integer $n$ and using $g(i) = \alpha^i$ for some $\alpha$), Rabin Fingerprint \cite{rabin-fingerprint} (working in $GF(2)$), and Buzhash (working with cyclotomic polynomials).

The reason rolling hashes are relevant to our topic is that many file backup services use variations of rolling hashes to achieve CDC. This paper will  primarily look at Tarsnap \cite{tarsnap}, a project by the second author, but we will also look at other schemes such as Borg \cite{borg} and Restic \cite{restic}.  Similar attacks may be possible on systems making use of rsync or related tools along with \texttt{gzip {-}{-}rsyncable}, which resets the compression state based on a window checksum.

Attacks on tools using these algorithms will generally consist of two parts:

\begin{enumerate}
    \item \emph{Parameter Extraction Attacks} extract the chunking algorithm's parameters, thus turning the algorithm into a deterministic algorithm known to the attacker. These attacks are specific to the protocol and can be thought of as individual instances of decryption problems.
    \item \emph{Post-Parameter Attacks} can be performed on the user's data after the chunking algorithm is known completely (such as after a parameter extraction attack, or in a situation where the chunking algorithm has no private information to begin with). These attacks tend to be more general and depend less on the specific chunking algorithm.
\end{enumerate}

In Section~\ref{sec:parameter-extraction}, we showcase parameter extraction attacks on Tarsnap, Borg, and Restic, which should translate into practical attacks after accounting for border cases and minor implementation details. In Section~\ref{sec:post-parametrization}, we show post-parameter attacks that can theoretically happen to \textbf{any} CDC-based chunking service. 

\subsection{Independent work}

After completing the original work underlying this paper in 2023, but before announcing it publicly in 2025, the authors learned of similar results by Truong et al.\cite{truong}  While the attacks described here and there are somewhat related, the work of both groups was performed entirely independently of each other.

\section{Preliminaries}
\label{sec:prelims}

\subsection{Attack Model} 
\label{sec:attack-model}

The most powerful attacker is usually the file backup service itself. In our attack model, we assume by default that the attacker is the server (but see Subsection~\ref{subsection:traffic} for alternatives). We assume that the client software is open-source so it does not do anything obviously bad, such as sending the server its chunking parameters. The attacker's general goal is to find out what files the user has uploaded, either now or in the future. 

For most of our parameter-extraction attacks, we assume \emph{known-plaintext} (the server knows some files that the user happens to have and how those are split into chunks) or \emph{chosen-plaintext} (the server can trick the user into uploading particular files and see how those are split) attacks. For post-parameter-extraction attacks we have a mixture of these and \emph{passive} (the server does not need any user-specific knowledge or influence) attacks.

\subsection{Information Flow and Encryption}
\label{sec:flow}

\begin{figure}
    \centering
    \includegraphics[height=2in]{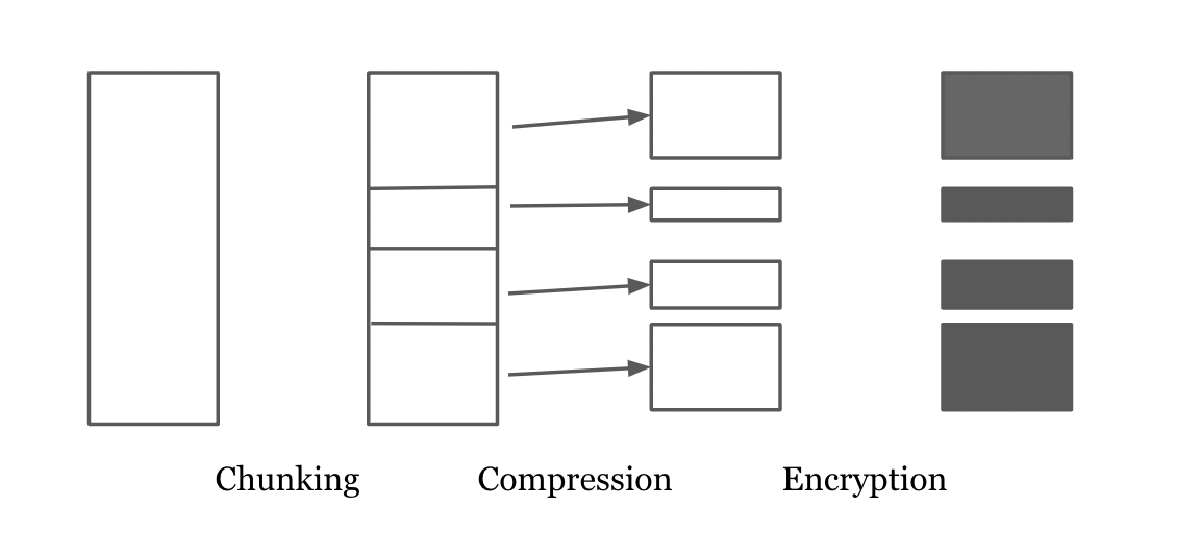}
    \caption{A visualization of data flow from the client to the server.}
    \label{fig:flow}
\end{figure}

The standard architecture taken by all of the services we consider (though different services may have different parts of this pipeline turned off and/or have different defaults) are:
\begin{enumerate}
    \item First, the data (say a file) is \textbf{chunked} using a known CDC scheme into strings $S_1, \ldots, S_n$.
    \item The strings are (sometimes optionally) \textbf{compressed} with some known compression algorithm $\chi$ into $\chi(S_1), \ldots, \chi(S_n)$.
    \item The individual pieces of compressed text $\chi(S_i)$ are \textbf{encrypted} with some industry-strength length-preserving encryption and then sent to the server.
\end{enumerate}
As a consequence, we assume that the server \textbf{knows the lengths of the compressed chunks} but not their content.  (Alternatively, even without access to the server, an attacker may be able to perform fine-grained traffic analysis to glean information about chunk sizes.  See Subsection~\ref{subsection:traffic} in the appendix for more details.)
We visualize the flow of data from the client to the server as in Figure~\ref{fig:flow}. 

In this architecture, the sequence of operations is basically ``forced'' by the situation:
\begin{enumerate}
\item Any modern (strong) encryption produces ciphertext which is effectively indistinguishable from random; consequently it is impossible to compress or usefully chunk the ciphertext if data is encrypted\footnote{We refer here, of course, to data being encrypted {\em at rest}.  Some services encrypt data {\em in transit}, e.g. using TLS, but not at rest; this allows them to deduplicate data between unrelated customers, but also introduces obvious privacy concerns.}.
\item In order for chunking to be useful, it must be possible to retrieve and decompress chunks individually; but this is only possible if the chunks are {\em compressed} individually, after the chunking has been performed.
\end{enumerate}

\subsection{Compression}
\label{sec:compression}
Compression can be thought of in a similar way across all our parameter-extraction attacks.

In our security model, we assume that the compression is with a fixed and known algorithm (such as the default one offered by the suite), which adds an overhead to our attacks. 
\begin{itemize}
\item In Tarsnap, the compression is done with zlib (compression level 9). 
\item Borg offers different compression options, of which lz4 is the default at the time of writing. 
\item Restic obtained compression in August 2022, and runs its own (open-source) algorithm.~\cite{restic-compression}
\end{itemize}

We call this algorithm $\chi(s)$. It creates some ambiguity when translating compressed chunk lengths $\lvert \chi(s)\rvert$ to uncompressed chunk lengths $\lvert s\rvert$. In effect, for \textbf{known or chosen plaintext attacks}, if we encounter a chunk with some compressed length $\lvert \chi(s)\rvert$ and $s$ is known to be a prefix of a known string, we can simply try the relatively few possible values of $s$.  We use the constant $k$ to denote the number of different possible values, or equivalently the compression factor.  For attacks which rely on highly-compressible chosen plaintexts, $k$ may be between $4$ and $8$, while for more generic inputs, the compression rate will typically be lower (which benefits the attacker since there is less ambiguity).  More specifically, for chosen plaintext attacks involving blocks of data consisting of only two distinct byte values, $k$ will approach the theoretical limit of $log(256) / log(2) = 8$; while for known plaintexts consisting of genomic data, the compression rates can range from the expected $log(256) / log(4) = 4$ to a more realistic $5$ (since base pairs in DNA are not entirely randomly selected from the four options).

For our work, we also consider attacks on versions of the protocol without compression for two reasons. One: it provides insight into how much security is ``offered'' by compression against our attacks. Two: not having compression is a legitimate option that certain users might choose to enable (or have already enabled!), so attacks under such contexts are also practical to study.

\subsection{Random Chunking Model}

Recall that $R_C$ is some subset of $R$ that will give a clash (and thus a new chunk) when the rolling hash becomes a value inside it. For our probabilistic arguments about chunking, we assume that at each new character we get a chunk (not accounting for minimum or maximum chunk length settings) with probability $\lvert R_C\rvert/\lvert R\rvert$. In this light, the set $R_C$ it typically chosen only with its size $\lvert R_C\rvert$ in mind, as a method of controlling the rate of chunking.

For example, a popular choice is to define $R_C$ to be the set of elements that when written in binary starts with some particular constant $m$ number of $0$'s, similar to Bitcoin's proof-of-work. This would cause the average chunk length to be $2^m$.

For somewhat technical reasons which we discuss later, the Tarsnap chunking algorithm doesn't exactly fit this model, though it is still useful as motivation.

\section{Parameter Extraction Attacks}
\label{sec:parameter-extraction}

In this section, we show how one can extract parameters from commonly used in-production file backup systems. Recall that this means that the attacker is able to see the chunk lengths but not be able break an industry-grade encryption protocol initiated by the user's client. By default, we typically first assume that there is no compression and then consider compression's effects separately.

Our attacks are fairly diverse, but all of them take the following outline:

\begin{tcolorbox}[breakable]
\textbf{General Attack Method:}
        \label{attack:general}
            \hfill
        \begin{enumerate}
\item Identify the space of chunking parameters and the rolling hash ring $R$.
\item Collect clashes (with a known or chosen plaintext). Convert each clash to some mathematical equation in $R$. 
\item Search the solutions to the set of equations in some efficient manner.
\item For each potential set of parameters $P$, check it on some (different) clash and accept if the chunk sizes induced by $P$ give the right chunk sizes. We call this last step a \emph{clash consistency check} of $P$. If the sizes are correct, accept (to reduce false positives, this part of the algorithm can be amplified by e.g. checking consistency on multiple clashes, as is standard in Monte Carlo algorithms).
\end{enumerate}
\end{tcolorbox}

\subsection{Tarsnap}

\emph{Tarsnap} \cite{tarsnap} uses a version of the Rabin-Karp type of hash, although instead of waiting for a rolling hash to equal $0$ in some ring $R$ (in this case $R = \mathbb{Z}/p\mathbb{Z}$ for some prime $p$), it waits for the running sum (with no sliding window) to equal any of the previous running sums that are sufficiently far away. This is similar to looking at rolling hashes with many sliding window sizes simultaneously and seeing if any of them equal $0 \in R$, though the precise details are still different for technical reasons.

\begin{tcolorbox}[breakable]
\textbf{Tarsnap:}

Secret parameters (derived from the user's secret key):
\begin{itemize}
    \item $p$, a prime close to $2^{24}$; there are $17 \approx 2^4$ possible choices.
    \item $\alpha$, a residue modulo $p$ of sufficiently high multiplicative order. There are roughly $2^{24}$ possible residues.
    \item $x[]: \{0, \ldots, 255\} \rightarrow \mathbb{Z}/p\mathbb{Z}$, a coefficient map that assigns to each possible byte some residue mod $p$. This is effectively a choice of $256$ random integers mod $p$.
\end{itemize}
Total parameter space size: $2^4 \cdot 2^{24} \cdot (2^{24})^{256} \approx 2^{6172}$.

\medskip

Chunking procedure: 
\begin{enumerate}
    \item Given any stream $S= b_1 b_2\cdots $ of bytes, at any place $J$ compute the ``running'' (\textbf{not} ``rolling'' since there is no window) hash $$y_J(S) = \alpha^1 x[b_1] + \alpha^2 x[b_2] + \cdots + \alpha^{J} x[b_J] \pmod{p}.$$
    \item We look for matches of this value to an earlier one.  If $y_K(S) = y_J(S)$ where $1 \leq K < J$, then we split at $J$, producing a chunk consisting of bytes from $1$ to $J$.
    \item We also always create a chunk if $J$ hits the maximal chunk size $261120 = (2^8-1)\cdot 2^{10}$.
\end{enumerate}

\end{tcolorbox}

There are further caveats in the actual implementation, such as requiring the inequality $J-K-1 < \lfloor \sqrt{4 J-\mu} \rfloor$ for the constant $\mu=2^{16}$ (corresponding to the mean chunk size) in order to chunk\footnote{The purpose of this constraint is to produce a faster-than-exponential decay in chunk size probabilities, that is to get more ``medium-sized'' chunks.} or zlib being used to compress the data. Going forward, we ignore these details (for Tarsnap but also other services) unless necessary in context.

Naively, we would have to try all $2^{6172}$ parameters combinations\footnote{Or the $2^{256}$ possible HMAC keys from which these parameters are generated.} on some generic known plaintext and see if any pass a chunking consistency check. We might have to repeat our process if the chunking consistency check is unreliable (that is, if many possible parameter sets would give the same chunk sizes). This is secure enough for practical purposes.

In fact, we were unable to find any known-plaintext parameter extraction attacks on Tarsnap that rely on a totally arbitrary file. However, we can make some progress if we assume a chosen plaintext model or if the known plaintext happens to have some desirable properties (which do occasionally occur in the wild).

The main idea of our attack is to try to break the coefficient map $x[]$ one at a time using \emph{small-alphabet} files, meaning files with few distinct byte values. Suppose we generate a known small-alphabet file $s$ that consists of only two distinct byte values, which we call $0$ and $1$ without loss of generality. This means that the chunking locations within the file $s$ depend on only four parameters: $p, \alpha, x[0], x[1]$. Furthermore, we can make our first reduction: observe that replacing $x[0]$ by $1$ and $x[1]$ by $x[1]/x[0]$ gives the same chunking behavior; this means we can effectively just assume that $x[0] = 1$ and look for $x[1]$ instead. At this point, a very straightforward attack is possible: 

\begin{tcolorbox}[breakable]
    \begin{attack}[Tarsnap, naive chosen-plaintext attack]
    \label{attack:tarsnap-naive}
    \hfill
        \begin{enumerate}
  \item Obtain $255$ clashes $C_1, \ldots, C_{255}$, with each $C_i$ from a chosen plaintext using only $0$ and $i$ as bytes.
   \item Enumerate all possible $(p, \alpha, x[1])$ triples.
   \item Perform a clash consistency check of $(p, \alpha, x[0] = 1, x[1])$ on $C_1$.
   \item Repeat a shortened version for all $i=2$ through $255$. For each $i$:
\begin{enumerate}
    \item Loop over all potential $x[i]$ (at this step, $p$ and $\alpha$ are already known).
    \item Perform a clash consistency check of $(p, \alpha, 1, x[i])$ on $C_i$. Accept if consistent.
\end{enumerate}   
        \end{enumerate}
    \end{attack}
\end{tcolorbox}

How long does this ``brute force'' attack take? For $b[1]$, the number of possible parameters is approximately
$$2^4 \times 2^{24} \times 2^{24} \approx 5\cdot 10^{15} \approx 2^{52}.$$ Suppose our known file is one megabyte in size (this is longer than necessary; see Section~\ref{sec:known-chunked} for discussion later). A naive estimate of chunking the file according to all possible parameters -- but not compressing them -- is on the order of $10^{10}$ CPU-hours ($100\,000$ CPU-years).  The task is ``embarrassingly parallel,'' so the clock time can be shortened by running on many cores.  While this attack is perhaps impractical, let us quickly upper-bound its cost: (cloud) compute can cost approximately 1 cent per 1 CPU-hour, so by another quite literal measure of cost, this is $\$100$ million of compute. Luckily, we do not have to repeat this workload $256$ times; after $p$ and $\alpha$ are known, the workload of the simplified loops are much smaller.

This naive attack is still not viable, although we were doing much better than $2^{6172}$ now. First, we reduced our search by a factor of $17$ million with the ``dividing out by $x[0]$'' strategy. Second, our chosen-plaintext attack allows us to ``divide and conquer'' and reduce most of the work to just finding~$\frac{x[1]}{x[0]}$. We now show that this direction can be made practical if we add one more ingredient: using the mathematical properties of the clashes to reduce the search space.

Recall that if there is a clash at position $J$ (recall that this means there is a chunk after position $J$ that is not caused by having hit the maximum chunk size), then we have the relationship $\sum_{i=0}^{J-1} x[s_i] \alpha^i \equiv \sum_{i=0}^{K-1} x[s_i] \alpha^i \pmod p$, which may be simplified to $\sum_{i=K}^{J-1} x[s_i] \alpha^i \equiv 0 \pmod p$ and further to $$\sum_{i=0}^{d-1} x[s_{J-d+i}] \alpha^i \equiv 0 \pmod p$$ where $d=J-K$.  In such a situation, let us call the substring $s_K s_{K+1}\cdots s_{J}$ that caused the hash collision a \emph{clash string} (of length $d$).

Suppose we knew the first uncompressed chunk size $J$ exactly. Assuming that this chunk came from a clash, we have a constraint relating $\alpha$ and $x[1]$ modulo $p$.  Specifically, we have $\sum_{i=J-d}^{J-1} x[s_i] \alpha^i \equiv 0 \pmod p$ for some clash length $d$. For each specific value of $d$, we have an explicit polynomial relationship for $\alpha$ and $x[1]$ roughly of the form $\sum_i x[s_i] \alpha^i \equiv 0 \pmod p$, where we elide the range of the $i$ and simplify some exponents for the sake of exposition.  We can now separate this as $\sum_{s_i=0} x[0] \alpha^i + \sum_{s_i=1} x[1] \alpha^i \equiv 0 \pmod p$, or rewriting,
\begin{equation}
\label{eqn:x1}
   x[1]/x[0] \equiv x[1] \equiv - \frac{ \sum_{s_i=0} \alpha^i} {\sum_{s_i=1} \alpha^i} \pmod p.
\end{equation}

In particular, if we knew the values $J$ and $d$ (and of course the string $s$), then $\alpha$ determines $\frac{x[1]}{x[0]}$ uniquely.  This allows the following attack:

\begin{tcolorbox}[breakable]
    \begin{attack}[Tarsnap, using clashes]
    \label{attack:tarsnap-clashes}
        \hfill
        \begin{enumerate} 
\item Input: $255$ long strings with the $i$-th string having only $0$ and $i$ as bytes for $i \in \{1, \ldots, 255\}$.
\item From the $i$th string, obtain two clashes $C_i$ and $C'_i$.  This gives $510$ clashes $C_1, \ldots, C_{255}$ and $C'_1, \ldots, C'_{255}$, with $C_i$ and $C'_i$ using only $0$ and $i$ as bytes.
\item Enumerate all possible $(p, \alpha, d)$ triples.
\item Compute the uniquely determined value of $x[1]$ as in Equation~\ref{eqn:x1} for both clash $C_1$ and clash $C'_1$.
\item Determine $(p, \alpha, x[1])$ by finding the overlap in the two lists obtained in the previous step.
\item Repeat a shortened version for all $i=2$ through $255$. For each $i$:
\begin{enumerate}
    \item For each $d$ (at this step, $p$ and $\alpha$ are already known), compute the uniquely determined $x[i]$ from the clash $C_i$.
    \item Perform a clash consistency check of $(p, \alpha, 1, x[i])$ on $C'_i$. Accept if consistent.
\end{enumerate}
\end{enumerate}
    \end{attack}
\end{tcolorbox}

On average, the constraint on $d$ allows for $400$ possible values\footnote{This estimation comes from the square root condition mentioned in the Tarsnap specs.}. This means we are looping over $2^4 \cdot 2^{24} \cdot 2^9 \approx 2^{37}$ sets of parameters; each parameter has an amortized cost of only a few operations since each of the sums in Equation~\ref{eqn:x1} can be reused in computing the sums for the next higher value of $d$.  Compression makes the situation a bit harder for the attack, but not overwhelmingly; on average for a file containing two distinct byte values, knowing the compressed chunk size allows approximately $k = 8$ possible uncompressed chunk sizes and thus $8$ possible {\em ending} points in addition to the $\approx 400$ possible {\em starting} points.  This results in an attack taking a few hundred CPU-hours; but the same considerations about parallelism from before apply here as well, so the attack can complete in under an hour given a few hundred CPU cores.

\subsection{Borg}

\emph{Borg} \cite{borg} uses Buzhash in "buzhash chunker" mode; it also provides a non-CDC "fixed chunker" mode that seems directly inspired by protection against the types of attacks we mention\footnote{From \cite{borg}: ``The buzhash table is altered by XORing it with a seed randomly generated once for the repository, and stored encrypted in the keyfile. This is to prevent chunk size based fingerprinting attacks on your encrypted repo contents (to guess what files you have based on a specific set of chunk sizes).''}. 
\begin{tcolorbox}[breakable]
\textbf{Simplified Buzhash Chunker}:

Secret parameters (derived from the user's secret key):
\begin{itemize}
    \item $x[]: \{0, \ldots, 255\} \rightarrow \{0, 1\}^{32}$, a coefficient map that assigns to each possible byte a $32$-bit string.
\end{itemize}
Total parameter space size: $(2^{32})^{255} = 2^{8160}$.

\medskip

Chunking procedure: 
\begin{enumerate}
    \item Given any stream $S= b_1 b_2\cdots $ of bytes, at any place $J$ compute the rolling hash $$y_J(S) = s^{N-1}(x[b_{J-N+1}]) \oplus s^{N-2}(x[b_{J-N+2}]) \oplus \cdots \oplus x[b_J] \in \{0, 1\}^{32},$$
    where $s$ is the cyclic bit shift function and $\oplus$ denotes XOR. $N = 4095$ is the hard-coded  window length.
    \item If the last $21$ bits of this hash equals $0$, then we chunk at $J$.
    \item We also chunk if we hit the maximal chunk size of $2^{23}$ bytes. 
\end{enumerate}
\end{tcolorbox}

Borg uses a $32$ bit Buzhash, which runs on XOR. At first glance (this will change by Attack~\ref{attack:borg-polynomials}) this means that we do not have the same algebraic properties as Tarsnap (in particular, no multiplication symmetry and no particular meaning to quantities like $x[1]/x[0]$).  However, since the window sizes are fixed, our search space is smaller compared to Tarsnap in this direction. In addition, we can make the problem ``linear'' in a way that removes the need for a ``divide and conquer'' approach.

To start this attack, for each $i \in \{0, \ldots, 255\}$, define $y_{i, j} \in \mathbb{Z}/2\mathbb{Z}$ for $j \in \{0, \ldots, 31\}$ to be bits such that $x[i] = y_{i, 0} y_{i,1} \cdots y_{i, 31}$. Our task is to solve for the $256$ variables $x[i]$ by solving for the $256\cdot 32 = 2^{13}$ variables $y_{i,j}$. Analogous to Tarsnap, a clash after $b_1 \cdots b_{4095}$ corresponds to some relation
$$ \bigoplus_{i=0}^{4094} s^i(x[b_i]) = d_0 \ldots d_{10} \underbrace{0 \cdots 0}_\text{21 zeroes}$$
where the $d_0$ through $d_{10}$ are binary digits.
\begin{tcolorbox}[breakable]
    \begin{attack}[Borg, linear algebra]
    \label{attack:borg-linear-algebra}
    \hfill
\begin{enumerate}
\item Input any (reasonably expressive) data; obtain $391$ clashes $C_1, \ldots, C_{391}.$
\item Each such relation corresponds to $21$ equations corresponding to one of the last $21$ bits in the XOR equaling $0$. These would all take the form 
$$y_{i_0, j} + y_{i_1, j+1} + \cdots + y_{i_{4094}, j+ 4094} = 0 \pmod{2},$$
where the second index in the subscripts is taken modulo $32$.
\item Solve the resulting linear system with the first $390$ clashes, which has $21\cdot 390$ equations and $2^{13}$ variables.
\item Perform a clash consistency check with the results on $C_{391}$.
\end{enumerate}        
    \end{attack}
\end{tcolorbox}

Because $21\cdot 391 > 2^{13}$, we have enough information to solve for all the variables (generically). This amount of data corresponds to about $21\cdot 2^{21}$ bits, or about $40$ megabytes. In an actual attack we would need actual clashes instead of chunks, which would account for some additional overhead. 

Attack~\ref{attack:borg-linear-algebra} will not work if compression is enabled, since any ambiguity factor in compression would be multiplicative across the $c$ clashes. We give a chosen plaintext attack for this case. First, we make the following algebraic observations, changing our bitstring operations to algebra over $R = GF(2)[X]/\langle X^{32}-1\rangle$:
\begin{enumerate}
    \item For each $i \in \{0, 1, \ldots, 511\}$, interpret $x[i]$ as a degree $\leq 31$ polynomial $P_i(X)$ over $GF(2)$; specifically, we can define a map $p: \{0, 1\}^{32} \rightarrow R$ such that $$ p(y_{i,0}y_{i,1}\cdots y_{i, 31}) = y_{i, 0}X^{31} + y_{i, 1}X^{30} + \cdots y_{i, 31}.$$
    \item The action of applying $s$ in the bitstring interpretation is equivalent to multiplying by $X$ modulo $X^{32} - 1$ in the polynomial interpretation. That is, $p(s(x[y])) = X p(x[y])$ in $R$.
    \item The action of XORing in the bitstring interpretation is equivalent to addition in the polynomial interpretation. 
\end{enumerate}

Now, we start by choosing only two types of bytes, say $0$ and $1$, in our chosen plaintext attack. Recall that each clash then corresponds to 
$$ \bigoplus_{i=0}^{4094} s^i(x[b_i]) = d_0 \ldots d_{10} \underbrace{0 \cdots 0}_\text{21 zeroes}$$
where each $b_i$ equals $0$ or $1$. Translated into $R$, this means
$$ \sum_{i=0}^{4094} X^i P_{b_i}(X) = \alpha_{31} X^{31} + \alpha_{32} X^{30} + \cdots + \alpha_{21} X^{21}$$
where we have two polynomials $P_0(X)$ and $P_1(X)$. Define $P_1'(X) = P_0(X) - P_1(X)$. We can rewrite the left-hand side as
$$ \left( \sum_{i=0}^{4094} X^i P_0(X) \right) + Q(X) P_1'(X),$$
where $Q(X)$ has degree at most $4094$ and accounts for the places where the byte is $1$ instead of $0$. 

By symmetry, $\oplus_{i=0}^{4095} s^i(x[j]) = 0\cdots 0$ (the upper summation limit is deliberate). This corresponds to $\sum_{i=0}^{4095} X^i = 0 \in GF(2)[X]$. Manipulating, we obtain that $\sum_{i=0}^{4094} X^i = X^{4095}  = X^{31}$ in $R$. Putting this together with the earlier constraint, we obtain
$$ X^{31} P_0(X) + Q(X) P_1'(X) = \alpha_{31} X^{31} + \alpha_{32} X^{30} + \cdots + \alpha_{21} X^{21}.$$
This means if we have $2$ clashes, we would be able to subtract off the $X^{31} P_0(X)$ term to obtain
$$ Q(X) P_1'(X) = \alpha_{31} X^{31} + \alpha_{32} X^{30} + \cdots + \alpha_{21} X^{21}.$$
In this equation, the $Q$'s and $\alpha$'s are the differences between the coefficients in the two constraints, where $Q$ is known to us and the $\alpha$'s are not. This means we can use a $2^{10}$-iteration search to solve for $P_1'(X)$. 

We are now ready to present our attack:
\begin{tcolorbox}[breakable]
    \begin{attack}[Borg, polynomials]
    \label{attack:borg-polynomials}
    \hfill
\begin{enumerate}
\item Input: $255$ long strings with the $i$-th string having only $0$ and $i$ as bytes for $i \in \{1, \ldots, 255\}$.
\item Using the first random string, define $P_1'(X) = P_1(X) - P_0(X)$. Obtain $4$ clashes.
\begin{enumerate}
    \item With the first $2$ clashes, obtain a constraint $$ Q(X) P_1'(X) = \alpha_{31} X^{31} + \alpha_{32} X^{30} + \cdots + \alpha_{21} X^{21}.$$ Loop over the $\alpha$'s to solve for $P_1'(X)$;
    \item Using the $3$rd clash, obtain a constraint $$ X^{31} P_0(X) + Q'(X) P_1'(X) = \beta_{31} X^{31} + \beta_{32} X^{30} + \cdots + \beta_{21} X^{21}.$$ Loop over the $\beta$'s to solve for $P_0(X)$.
    \item Using the $4$th clash, perform a clash consistency check on the proposed $P_0(X)$ and $P_1'(X)$ (equivalently, $P_0(X)$ and $P_1(X)$).
\end{enumerate} 
\item Now that we have $P_0(X)$, repeat a shortened version of the above for the other $254$ long strings. On string $i \geq 2$, define $P_i'(X) = P_i(X) - P_0(X)$. Obtain $2$ clashes.
\begin{enumerate}
    \item With the $1$st clash, obtain a constraint $$ X^{31} P_0(X) + Q(X) P_i'(X) = \alpha_{31} X^{31} + \alpha_{32} X^{30} + \cdots + \alpha_{21} X^{21}.$$ Loop over the $\alpha$'s to solve for $P_i'(X)$;
    \item Using the $2$nd clash, perform a clash consistency check on $P_0(X)$ and $P_i'(X)$.
\end{enumerate} 
\end{enumerate}        
    \end{attack}
\end{tcolorbox}

This attack greatly reduces the multiplicative overhead of compression since we only need $4$ loops. For non-compressed setups, the other attack is more direct and more efficient.

\subsection{Restic}
\label{sec:restic}

Restic \cite{restic} uses straight Rabin fingerprint on $64$ bytes (= $512$ bits), chunking when the lowest $21$ bits of the resulting hash equals zero.

\begin{tcolorbox}[breakable]
\textbf{Restic:}

Secret parameters:  
\begin{itemize}
    \item $P[X]$: a random irreducible polynomial over $GF(2)$. This is generated once and is saved, encrypted, into a file in the repository.
\end{itemize}

Chunking procedure: 
\begin{enumerate}
    \item Given any stream of bytes $S$ ending in $64$ bytes $B_1 \ldots B_{64}$, convert the last $64$ bytes into $64\cdot 8 = 512$ bits $b_0 \ldots b_{511}$ and compute the rolling hash $$y(S) = b_0 X^{511} + b_1 X^{510} + \cdots + b_{510} X + b_{511} \in GF(2)[X],$$
    where the result is considered to be a polynomial in $X$ over $GF(2)$.
    \item If the result modulo $P[X]$, seen as a binary string, has its last $21$ bits equal to $0$, then we chunk at $J$.
    \item Also chunk if we hit the maximum chunk size of $2^{23}$ or $2^{30}$ bytes.
\end{enumerate}
\end{tcolorbox}

Unlike the other two systems, Restic does not have an obscured coefficient map $x[]$;  the entirety of the work is in breaking the hidden modulus $P[X]$, which is a degree $53$ polynomial over $GF(2)$. For each $i \in \{0, \ldots, 52\}$, define $y_{i} \in \mathbb{Z}/2\mathbb{Z}$ via the equation
    $$ P[X] = X^{53} + y_{52}X^{52} + \cdots + y_1 X + y_0.$$
Our task is to solve for $P[X]$ by  solving for the $53$ variables $y_{i}$. As before, we obtain our information through clashes. Each clash after a substring of $64$ bytes corresponds to a substring of $64\cdot 8 = 512$ bits $b_0 \cdots b_{511}$, which gives the equation
$$ \sum_{i=0}^{511} b_i X^{511-i} = Q(X) P(X) + R(X),$$
where $P(X)$ is our hidden degree $53$ polynomial, $Q(X)$ is a degree $458$ (roughly) quotient polynomial, and $R(X)$ is a degree $\leq 52$ remainder polynomial with the last $21$ terms equal to $0$.

We present two attacks for Restic. First, we present a linear-algebra based algorithm similar to Attack~\ref{attack:borg-linear-algebra} that is fast when there is no compression. Then, we present a different attack that involves polynomial GCDs which is even faster without compression, but also works with compression as there are fewer loops.
\begin{tcolorbox}[breakable]
\begin{attack}[Restic, solving $R(X)$]
\label{attack:restic-solve}.
\hfill
\begin{enumerate}
\item Input any (reasonably expressive) data. Obtain $459$ clashes.
\item Each clash gives a relation of the form $A_i(X) = P(X)Q_i(X) + R_i(X)$, where $A_i(X)$ has degree $511$ and $R_i(X)$ has degree 52 with 21 known zero bits.
\item Using Gaussian elimination, cancel off the higher-order terms in the $A_i(X)$ to obtain some $A'(X) = P(X) Q'(X) + R'(X)$ where $A'(X)$ has degree $53$ and 
$R'(X)$ has degree $52$ with the same $21$ known zero bits.
\item Based on the degrees we must have $Q'(X) = 1$, and thus the coefficients of $A'(X)$ and $P(X)$ match for the terms corresponding to the known zero bits in
$R'(X).$
\item Now that we have $21$ bits of $P(X)$, loop through the other $31$ bits.
\item For each choice of $P(X)$, perform a clash consistency check.
\end{enumerate}
\end{attack}    
\end{tcolorbox}

The $4$th step fails in the degenerate case where $A'(X) = 0$ because everything canceled, but this should only happen with probability $1/2$ per run, so it vanishes as we get more clashes. In general, we will take on the additive complexity of the $O(n^3)$ work from the Gaussian elimination and the $2^{31}$ operations from the search.

\begin{tcolorbox}[breakable]
\begin{attack}[Restic, guessing $R(X)$]
\label{attack:restic-guessing}
\hfill
\begin{enumerate}
\item Obtain $3$ clashes. For the first clash, this gives $512$ bits $b_0 \cdots b_{511},$ which gives a polynomial $\sum_{i=0}^{511} b_i X^{511-i} = S(X).$
\item Loop over the $2^{53-21} = 2^{32}$ possibilities of the first $32$ terms of $R(X)$. Each such $32$-tuple gives one purported value of $R(X)$.
\begin{enumerate}
    \item Compute the GCD\footnote{Recall that $X^{2^{53}-1}+1$ is the product of all irreducible polynomials in $F_2[X]$ of degree dividing $53$; this GCD can be rapidly computed via repeated squaring modulo $S(X)-R(X)$ followed by a GCD of two degree-$511$ polynomials.} of $S(X)-R(X)$ and $X^{2^{53}}+X$ mod $2$.
    \item Factor the result (if nontrivial) and enumerate irreducible divisors of degree $53$.
    \item Set $P(X)$ equal to any such divisors and perform a clash consistency checks on the remaining $2$ clashes.
\end{enumerate}
\end{enumerate}
\end{attack}    
\end{tcolorbox}

We know that there are $\frac{2^d-2}{d}$ irreducible polynomials of degree $d$, so the expected number of irreducible degree-$p$ factors of a random higher-degree polynomial is $\frac{2^d - 2}{d 2^d}$. This means in our $2^{32}$ searches, we should have found roughly $2^{26}/53 \approx 2^{26}$ irreducible polynomials of degree $53$; furthermore, in all but $1/(2 \cdot 53^2)$ of the $2^{32}$ searches the result of the GCD is degree $0$, $1$, $53$, or $54$, so the requisite polynomial factorizations are rare enough to not be significantly expensive.  For any particular ``false positive'' irreducible polynomial that's not actually $P(X)$, the probability that on a second clash it would have a remainder with the last $21$ terms equal to $0$ should be only about $2^{-21}$. This would still have room for about $2^{26 - 21} \approx 32$ false positives, which means we want a third clash to test for false positives.  Thus, this method should be good enough to find the correct $P(X)$.

Unlike with Borg, where different attacks have different strengths, Attack~\ref{attack:restic-guessing} should be faster than Attack~\ref{attack:restic-solve} regardless of the compression factor $k$.

\subsection{Summary}

We summarize our parameter extraction attacks in Table~\ref{tab:parameter-extraction-attacks}; for these attack times,
\begin{itemize}
    \item $k$ refers to the compression overhead for known-plaintext attacks (each compressed chunk size corresponds to $k$ possible uncompressed chunk sizes).
    \item We assume Gaussian elimination on $n$ pivots over $GF(2)$ takes $n^3/64$ operations.
    \item We assume computing the GCD of two polynomials of degree $n$ over $GF(2)$ takes $n^2/64$ operations.
\end{itemize}

\begin{table}
\begin{tabular}{ |p{1.5cm}||p{1.5cm}|p{3cm}|p{3cm}|p{3cm}|  }
 \hline
 \multicolumn{5}{|c|}{Summary of Attacks} \\
 \hline
 Attack & Service & Chosen or Known plaintext & compute required & number of chunks \\
 \hline
 Attack~\ref{attack:tarsnap-naive}  & Tarsnap & Chosen & $2^{52}$ & 255 \\
 \hline
 Attack~\ref{attack:tarsnap-clashes} & Tarsnap & Chosen & $2^{42}$ & 510 \\
 \hline
 Attack~\ref{attack:borg-linear-algebra} & Borg & Known & $2^{21} \cdot k^{389}$ & 391 \\
 \hline
 Attack~\ref{attack:borg-polynomials} & Borg & Chosen & $2^{42}$ & 512 \\
 \hline
 Attack~\ref{attack:restic-solve} & Restic & Known & $2^{52} \cdot k^{458}$ & 459 \\
 \hline
Attack~\ref{attack:restic-guessing} & Restic & Known & $2^{44} \cdot k$ & 3 \\
 \hline
\end{tabular}
\caption{\label{tab:parameter-extraction-attacks} Parameter extraction attacks. }
\end{table}

The attacks given here are tactical -- they are fairly distinct, though they share some algebraic similarities. The main unifying theme is that the hashes used for these systems leak a nontrivial amount of information, so the secrets can be recovered with a dedicated search that is optimized for each use case.

\subsection{Potential defences}

We now discuss a few ideas (by no means exhaustive) on how to protect against the attacks, with the common strategy being to enlarge the key space for the secret information used during chunking. There are also some system-specific considerations.

\begin{itemize}
    \item The Tarsnap attack we describe focuses on files with a small number of distinct byte values. A proposal for Tarsnap to defend against such attacks is to use a rolling hash that ``forces'' having many types of bytes. We propose that when considering a new byte $b_n$, instead of adding $\alpha^i x[b_n]$ to the running hash, add $\alpha^i x'[b'_n]$, where
$$ b_n' := H'(b_{n-k + 1} b_{n-k+2} \cdots b_n)$$
is computed with a second rolling hash (with different parameters) $H': \{0, \ldots, 255\}^k \rightarrow 2^{16}$ to provide an input to an enlarged coefficient map $x': 2^{16} \rightarrow 2^{256}$. This combination effectively turns any text (for chunking purposes) to one where very few bytes are repeated. Thus, all of our attacks (which depend on files with few repeating values) will fail. The cost is that the chunking becomes slower since we need to do an additional rolling hash per byte. Enlarging the prime space is also a conceptually simple way to add a few orders of magnitude of keyspace security.

\item In Borg, if one forces the option of not having compression, for the linear algebraic Attack~\ref{attack:borg-linear-algebra}, assuming that it takes $O(n^3)$ to solve a system of $n$ linear equations, the $n$ in our attack is $256\cdot 32 = 2^{13}$, so we are in the (up to a constant) $2^{39}$ range, which is similar to where we are in the Tarsnap attack. Naively, this means if we just made the Buzhash $8$ times bigger to $256$ bits instead, we would already get to $2^{48}$, which is starting to look much harder. However, like with the chosen plaintext attacks Attacks~\ref{attack:tarsnap-clashes} and ~\ref{attack:borg-polynomials}, the attacker might be able to use a chosen plaintext attack that attacks two byte values at once. This would then need to be repeated $256/2 = 128$ times, but each time instead of having $256\cdot 256$ variables, we would just have $256\cdot 2$ variables, which corresponds to (up to a constant) $2^{27}$ computations, which gives a total estimate of $2^{27}\cdot 128 = 2^{34}$ ballpark again. Thus, against a sophisticated attacker we would probably need to make the hashes about $2^{10}$ larger. Then even if only two byte values are attacked at once, it would take on the order of $2^{60}$ computations to extract the coefficients. 

If we only worry about the case of compression enabled, then the only attack we have available is Attack~\ref{attack:borg-polynomials}. In this case, simply changing hashes to having $64$-bit codomain instead of $32$-bit would change the $(2^{10})^2$ search in the algorithm to $(2^{42})^2$, which should help protect against the attack.

\item The Restic attacks are similar in spirit with the Borg attacks. A simple keysize extension would be enough to defend against our particular attacks. If the polynomial degree were to be doubled to $\approx 128$ or so, we would have to run through $\approx 2^{128 - 21} = 2^{107}$ iterations in our search loop, which offers more protection.

\end{itemize}

\section{Post-Parametrization Attacks}
\label{sec:post-parametrization}

We see two ``natural'' classes of attacks after a rolling-hash type CDC is parameterized completely:
\begin{enumerate}
    \item Known Chunked: the attacker only knows the chunk sizes and then tries to find out information about the plaintext being sent. 
    \item Partially Chosen Chunked: the attacker can manipulate the user to input secret information into a file (to be uploaded) where the other content can be controlled by the attacker.
\end{enumerate}

We were not able to find our exact attack scenarios in literature, but there was some precedent in adjacent fields. Kelsey \cite{kelsey-compression} describes very similar attacks in the context of compression algorithms. ``Stateless compression side-channel attacks'' were used to describe attacks where only the compression ratios are known, and "stateful compression side-channel attacks" were used to describe attacks where information about the rest of the message is known or controlled. These two types of attacks roughly correspond to our two classes of attacks respectively.

\subsection{Known Chunked Attack}
\label{sec:known-chunked}

We start by getting a relevant but extremal case out of the way:
\begin{ex} [Single Known File]
\label{ex:unchunked}
As an extremal but still relevant example, one can consider the situation that the attacker has a particular known file in mind and wants to detect whether the user has uploaded it. A backup server as the attacker (or someone who is able to sniff chunk sizes from the user-server communications) with any particular file in mind would be able to detect that the user has uploaded the file at any point.
\end{ex}

For deeper analysis of the known chunked situation, since there is nothing the attacker really does besides looking at the chunk sizes, the known chunked attack can be analyzed through the perspective of the information-theoretical ``leakage rate'' from chunking. That is, we assume some prior distribution of data (which by default, we assume is uniform). Then, we look at the distribution of chunk sizes and consider its entropy, which gives an approximation of how many bits of information are ``leaked'' per chunk.

\begin{figure}
    \centering
\includegraphics[height=2in]{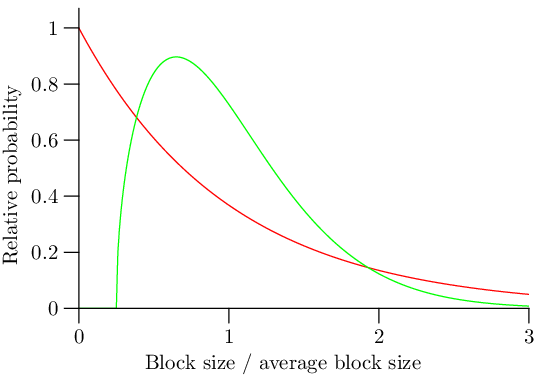}
    \caption{The geometric distribution drawn in red is the distribution of uncompressed chunk sizes for most chunking algorithms, such as in Borg or Restic. The unimodal distribution drawn in green is the distribution of uncompressed chunk sizes in Tarsnap.}
    \label{fig:tarsnap-distribution}
\end{figure}

\begin{itemize}
    \item To start, we look at Tarsnap. On average, an uncompressed Tarsnap chunk size is around the mean chunk size parameter $\mu = 2^{16} = 65\,536$ bytes; it is actually slightly larger, with the distribution as seen in Figure~\ref{fig:tarsnap-distribution}.
    The entropy of the uncompressed chunk size is $16$ or $17$ bits per chunk.  If the chunk is mostly uniformly random data, then the entropy of the compressed chunk size is similar.
    As a result, the information leakage is approximately $(16\text{ bits})/(2^{16}\text{ bytes}) = 2^{-15} \approx 1/{30\,000}.$
    \item Borg and Restic are very similar in that (as of time of this document) their min and max chunk sizes are both $2^{19}$ and $2^{23}$ respectively and they both use $1/2^{21}$ as roughly the probability of chunking at any given point. Assuming a random chunking model (inducing a bounded geometric distribution on the chunk sizes), this means the uncompressed chunk size distribution has roughly $21$ bits of information per chunk. The average chunk length is around $2^{21}$, so the information leakage is approximately $(21\text{ bits})/(2^{21}\text{ bytes}) \approx 2^{-17} \approx 1/{100\,000}.$

\end{itemize}

In other words, at least one hundred-thousandth of random data is leaked.  If the data in question is at least 1.6 megabytes, then this is at least $16\text{ bytes}=128\text{ bits}$.  In particular, for random data, this is ``universally uniquely identifiable'' in the sense of ``universally unique identifiers (UUIDs)'', as UUIDs are $128$ bits long. The main implications are as follows: if Alice is suspected to have one of a collection of known files, each of which is at least a couple megabytes in size and which are sufficiently ``random'' and different from each other, the file Alice possesses can be identified uniquely.

\begin{ex} [Music]
\label{ex:music}    
Using Tarsnap, we chunked a music library consisting of $10\,000$ songs each downloaded from a known, external repository.  The first two compressed chunk sizes uniquely identified the album for each song. For most songs, the first two (and certainly the first three) compressed chunk sizes uniquely identified the song.  However, for a handful of albums, the MP3 files from the same album began with the same data, usually because the files included the same album art for each file.

Even for a large library from iTunes or Spotify, similar numbers of chunks are expected to uniquely identify the songs.  This is because even $2^{2\cdot 16}\approx 4\text{ billion}$ (the implied number of different possibilities for two chunk lengths, information-theoretically) far exceeds the number of songs ever made.
\end{ex}

As Example~\ref{ex:music} shows, probably every single MP3 file that Alice possesses may be identified, if the files are taken from a known collection (e.g. iTunes or other official/centralized sources).  The same should be true for other media files, such as videos. In general, any large file that Alice is suspected of having may be verified using this attack.  This includes anything ``publicly available,'' such as any file from The Pirate Bay or Wikileaks.

\begin{ex}[DNA]
\label{ex:DNA}
The personal genomics company 23andMe \cite{23andme} offers a service that collects some of the user's SNPs (single nucleotide polymorphisms). These SNPs can be downloaded, producing a file that looks like the following:

\begin{verbatim}
    # This data file generated by 23andMe at: Mon Oct 13 19:07:27 2014
    #
    [...]
    # 
    # More information on reference human assembly build 37 (aka Annotation Release 104):
    # http://www.ncbi.nlm.nih.gov/mapview/map_search.cgi?taxid=9606
    #
    # rsid  chromosome      position        genotype
    rs12564807      1       734462  AA
    rs3131972       1       752721  AG
    rs148828841     1       760998  CC
    rs12124819      1       776546  AA
    rs115093905     1       787173  GG
    rs11240777      1       798959  AG
    rs7538305       1       824398  AC
    rs4970383       1       838555  CC
    rs4475691       1       846808  CC
    rs7537756       1       854250  AA
\end{verbatim}
    
One thing that Alice may do with this data is ``paste'' her genotypes into the reference genome, that is, replacing the base (nucleotide) at a given position with her own.  While this is a somewhat unusual operation for several reasons (for example, she has two copies of each chromosome, so it's unclear which of the two bases to paste -- 23andMe does not distinguish between the copies), we consulted with a geneticist and it appears this is in fact an operation that computational geneticists perform.  In particular, files like this can be found ``in the wild.'' 

Upon having done so, suppose Alice uploads this modified genome to Tarsnap. Suppose the file is encoded with one base per byte (which is typical), so that this is a small-alphabet file with $m=4$.  Notice that in the file provided above by 23andMe, the very first SNP occurs at position $734\,462$.   This is well over the half a megabyte of data necessary to determine the parameters $p,\alpha,b[\mathrm{A}],b[\mathrm{C}],b[\mathrm{T}],b[\mathrm{G}]$ with some of the attacks above.  In particular, even though Alice has uploaded a modified genome, a very large prefix of the data is identical to the reference genome, so it may as well be considered ``known.''

However, starting at the position of the first SNP above, Alice's genome potentially differs from the reference genome, so at that point, the attacker may attempt to determine Alice's genotype. There are a few ways in which the compressed chunk lengths from Alice's genome may leak information.

First, the differences between Alice's genome and the reference may change the breakpoints. This occurs sometimes (we estimate that one-sixth of the split points of a chunk might be different in one of these genetic examples), and can leak a large amount of information, usually enough to solve for the changed data completely. For example, if some string $(S_1||S_2||S_3)$ did not chunk but $(S_1||S_2'||S_3)$ now chunks after $S_3$, where $S_2'$ is Alice's hidden data, we can probably immediately solve for $S_2'$ assuming reasonable bounds on the length of $S_2'$, especially if the alphabet is severely limited to e.g. $A, C, T, G$.

Alternatively, suppose that the breakpoints are identical and the genotype does not involve an insertion or deletion (relatively uncommon in 23andMe data: approximately half a percent of all SNPs). Then the uncompressed chunk lengths are the same in Alice's genome and the reference genome.  However, the information transmitted over the wire leaks the compressed chunk lengths, and those also depend on the data!
\begin{enumerate}
    \item For example, if the SNPs are relatively far apart and only a single SNP falls within a chunk, it may be the case that the compressed chunk length would depend on the SNP as follows: $A:7051$, $C:7050$, $T:7051$, $G:7050$. As such, the attacker is able to determine partial information about Alice's genome.  Here, assuming the different bases are equally likely, one bit was leaked from the compressed chunk length, half of the total of two bits per base. In reality, there is a prior distribution on bases that may leak further information. For example, if only $A$ and $C$ are seen in the wild, then the length would completely determine the SNP.
\item Extending, we can have multiple bases fall within a single chunk, in which case the information is more diffuse but can still be analyzed. Suppose five bases are chunked together.  Then the distribution of compressed chunk lengths may be as in Table~\ref{tab:distribution_lengths}. In this example, on average $\approx 2.6$ bits out of a total $10$ bits would be leaked from this chunk (again assuming the different $5$-tuples were equally likely).  But if the length were $19422$, the full $5$-tuple would be known.
\end{enumerate}
We estimate (very crudely!) that $20$\% of the information content of a 23andMe data file may be leaked in this manner, primarily due to the change in chunking but also partially from the differences in compressed length.

\begin{table}
    \centering
    \begin{tabular}{|r|r|r|r|r|r|r|r|r|r|}
    \hline
Length & 19414 & 19415 & 19416 & 19417 & 19418 & 19419 & 19420 & 19421 & 19422 \\ \hline
Count & 16 & 182 & 336 & 192 & 53 & 101 & 108 & 35 & 1 \\ \hline
    \end{tabular}
    \caption{An example distribution length of number of $5$-tuples that result in each particular length of a compressed chunk. The first row is the length of the compressed chunk containing these particular 5 nucleotides, and the second row is the number of $5$-tuples resulting in this length.}
    \label{tab:distribution_lengths}
\end{table}
\end{ex}

Example~\ref{ex:DNA} is a real-world example of a not-quite-known\footnote{What we mean by this is that the plaintext is technically a "chosen" plaintext, but the situation occurs in the wild, so it may occur without the attacker specifically \emph{choosing} this plaintext.} plaintext situation with relatively small $m=4$, that leaks actual information of interest that people are concerned with keeping secure. It also demonstrates a curious mechanism of information leakage.  Because SNPs are relatively rare, most of them do not participate in a clash and the chunks are mostly driven by the reference genome.  Thus, in those cases, the uncompressed chunk lengths wouldn't have leaked information; instead, the compression reveals information here!

\subsection{Partially Chosen Chunked}
\label{sec:partially-chosen-chunked}

Some examples of the partially chosen chunked setup are:
\begin{itemize}
    \item The attacker (possibly the owner of the database server) can trick the user to fill out some items in the a database with some secret, with the resulting database being exported and backed up. But the attacker is able to control the text surrounding the item (such as adjacent entries, strings encoding the ``structure'' of the database, or comments).
    \item The attacker can trick the user to fill out and backup a PDF survey or form, with ``inactive'' text adversarially edited before and after the entry.
    \item (inspired by Example~\ref{ex:DNA}) The attacker can trick the user to inject some of their personal genome into a purported ``standard'' genome file, but adversarially edit some of the ``inactive'' bases around the personalized parts in preparation for this attack.
\end{itemize}

\begin{tcolorbox}[breakable]
\begin{attack}[Standard Partially-Chosen Chunked Attack]
\label{attack:partially-chosen-attack}
\hfill
\medskip
Output: a ``partially completed'' string of the form $(p||[c]||s)$, where:
\begin{enumerate}
\item $p$ is a prefix known to the attacker;
\item $[c] \in C$ is a secret input (of fixed known size) known only to the user and constrained to some set $C$, to be filled by the user later.
\item $s$ is a suffix controlled by the attacker.
\end{enumerate}
\medskip
Attack setup:
\begin{enumerate}
    \item Start with $p$ as given and $s$ equal to the empty string. We now construct $s$ one character $s_i$ at a time after the $[c]$ area.
    \item Emulate the state of the chunking scheme at this position; that is, for each possible $c \in C$, compute the current rolling hash value in $R$.
    \item For each of the possible $\lvert B\rvert$ values of $s_i$, say that $s_i$ \emph{clashes assuming} $c_j$ if $(p||c_j||s_1\cdots s_i)$ creates a clash.
        \begin{enumerate}
            \item If writing a particular $s_i$ would clash assuming some $c_j$, append $s_i$ to $s$, remove $c_j$ from $C$, and continue.
            \item If not, then pick a random $s_i \in B$ to append to $s$. 
        \end{enumerate}
    \item Update the state of the chunking scheme for the remaining elements in $C$. Continue until $\lvert C\rvert = 1$.
\end{enumerate}
\end{attack}
\end{tcolorbox}

As the probability of clashing is $\lvert R_C\rvert/\lvert R\rvert$ in the random chunking model, the probability of our state having a $s_i$ that would clash assuming some $c_j$ is roughly $p = \lvert B\rvert\cdot \lvert R_C\rvert/\lvert R\rvert$. This means at each step, we have found a $s_i$ that would ``identify'' one of the $\lvert C\rvert$ choices uniquely with probability $p$, and otherwise get reset into the same state. Treating this as a simple Markov chain, the expected number of characters before we are able to remove one of the choices of $C$ is $1/p$. By linearity of expectations, the average length of $s$ we would need to chunk all but one possibility is $$\frac{\lvert C\rvert}{p} = \frac{\lvert C\rvert\cdot\lvert R\rvert}{\lvert B\rvert\cdot\lvert R_C\rvert}$$
(strictly speaking, we have $\lvert C\vert-1$ instead of $\lvert C\rvert$, since once we go from $2$ options to $1$ option of $C$ we have differentiated all options). We now have a string that chunks at a different chunk length for each possible input $c \in C$.

A useful way of conceptualizing this attack is that it speeds up the information leakage rate (as described in Section~\ref{sec:known-chunked}) by $\lvert B\rvert$ compared to a passive known chunked attack, since the attacker gets to choose to look at $\lvert B\rvert$ characters at a time instead of leaving it up to chance. Also, the ``prefix'' / ``suffix'' distinction is not really important; if the attacker has access to both the prefix and the suffix, they can attack by changing both, at a rate of $\lvert B\rvert$ per character in the prefix or suffix.

\begin{ex}[Citizenship / binary choices]
As a proof of concept, we can assume the attacker wants to figure out if the user is a U.S. citizen by inducing the user to sign some citizenship form with a ``yes/no'' option. Then if the form is saved in a file with the format $(p||c||s)$ as given and we are streaming bytes with chunking probability $1/2^{21}$ (as in Borg or Restic), it would take an average of $2^{21}/2^8 = 2^{13}$ bytes to find a difference.     

For Tarsnap in particular, the chunking probability changes with $J$ because the behavior of Tarsnap's chunking algorithm is similar to "doing many rolling hashes at once." When $J$ gets sufficiently close to the maximum chunking length $261120$, the Tarsnap specs give roughly $\sqrt{4\cdot 261120-65536} \approx 2^{10}$ potential windows, a rolling hash equaling to $0$ of any of them creating a chunk. Since $p \approx 2^{24}$, this means the chunking probability is roughly $2^{14}$ at that point, which means it would only take $2^{14}/2^8 = 2^{6}$ bytes to find a difference if the attacker "aims" to have $\lvert(p||c||s)\rvert$ get close to $261120$.
\end{ex}

\section{Conclusion}

\subsection{Parameter-extraction Attacks}

The main theme of our parameter-extraction work is that \textbf{CDCs are essentially using computationally efficient (instead of cryptographically secure) hash functions as a sort of pseudo-encryption, with the chunking parameters acting as a secret key.} This means that they are susceptible to some attacks and their security-efficiency tradeoffs should be strongly considered. 

It seems like compression as default (or even required) is important. Without compression, Borg and Restic are susceptible to known plaintext attacks. With compression, we still have theoretically sound (and harder) chosen-plaintext attacks but no known-plaintext attacks. Sadly, compression can also leak information for post-parameter extraction attacks, as shown in Example~\ref{ex:DNA}.

\subsection{Post-parameter-extraction Attacks}

Every chunking algorithm leaks some basic rate $r$ (for Tarsnap, $1/30000$) of information assuming uniformly random data. Example~\ref{ex:music} already demonstrated how this can be used in practice to learn something concrete. However, in specific real-world situations, the leakage rate can be much higher than our ``maximum entropy'' assumption of uniformly random data:
\begin{enumerate}
    \item In a low-entropy setting (such as DNA or surveys where there are small changes from a known ``default'' data), the user changes can cause chunking breakpoint changes that give away the information completely.
    \item Even if the chunking breakpoints do not change, the chunk sizes from compression would give away significant information from the data, as in Example~\ref{ex:DNA}. See \cite{kelsey-compression} for related work.
    \item When an attacker can influence the ``default'' data (such as in the case of a malicious survey in Section~\ref{sec:partially-chosen-chunked}), the information can be adversarially extracted faster than known-chunking attacks by roughly a factor of $\lvert B\rvert$, the size of the alphabet.
\end{enumerate}

\bibliographystyle{acm}
\bibliography{bibliography}

\begin{thebibliography}{10}

\bibitem{23andme}
{\sc {23andMe}}.
\newblock 23andme: Personal genomics and biotechnology company.
\newblock \url{https://www.23andme.com/}.
\newblock Accessed: 2023-05-29.

\bibitem{borg}
{\sc {Borg}}.
\newblock Borg -- deduplicating archiver with compression and encryption.
\newblock \url{https://www.borgbackup.org/}.
\newblock Accessed: 2023-05-29.

\bibitem{rabin-karp}
{\sc Karp, R.~M., and Rabin, M.~O.}
\newblock Efficient randomized pattern-matching algorithms.
\newblock {\em IBM Journal of Research and Development 31}, 2 (1987), 249--260.

\bibitem{kelsey-compression}
{\sc Kelsey, J.}
\newblock Compression and information leakage of plaintext.
\newblock In {\em Fast Software Encryption\/} (Berlin, Heidelberg, 2002),
  J.~Daemen and V.~Rijmen, Eds., Springer Berlin Heidelberg, pp.~263--276.

\bibitem{rabin-fingerprint}
{\sc Rabin, M.~O.}
\newblock Fingerprinting by random polynomials.
\newblock Tech. Rep. TR-15-81, Center for Research in Computing Technology,
  Harvard University, 1981.

\bibitem{restic}
{\sc {Restic}}.
\newblock Foundation - introducing content defined chunking ({CDC}).
\newblock \url{https://restic.net/blog/2015-09-12/restic-foundation1-cdc/},
  2015.
\newblock Accessed: 2023-05-29.

\bibitem{restic-compression}
{\sc {Restic}}.
\newblock restic 0.14.0.
\newblock \url{https://github.com/restic/restic/releases/v0.14.0}, August 2022.

\bibitem{RFC879}
{The TCP Maximum Segment Size and Related Topics}.
\newblock RFC 879, Nov. 1983.

\bibitem{tarsnap}
{\sc {Tarsnap}}.
\newblock Tarsnap -- online backups for the truly paranoid.
\newblock \url{https://www.tarsnap.com/}, 2015.
\newblock Accessed: 2023-05-29.

\bibitem{truong}
{\sc Truong, K.~T., Merz, S.-P., Scarlata, M., G{\"u}nther, F., and Paterson,
  K.~G.}
\newblock Breaking and fixing content-defined chunking.
\newblock Unpublished manuscript (submitted), January 2025.

\bibitem{Williams1999Partitioning}
{\sc Williams, R.~N.}
\newblock Method for partitioning a block of data into subblocks and for
  storing and communicating such subblocks, November 23 1999.
\newblock U.S. Patent 5,990,810.

\bibitem{Xia2016Comprehensive}
{\sc Xia, W., Jiang, H., Feng, D., Douglis, F., Shilane, P., Hua, Y., Fu, M.,
  Hu, Y., Zhang, Y., Zhou, Y., and Zhao, M.}
\newblock A comprehensive study of the past, present, and future of data
  deduplication.
\newblock {\em Proceedings of the IEEE 104}, 9 (2016), 1681--1710.

\end{thebibliography}

\appendix

\section{Chunk discovery}

In order to carry out any of the attacks mentioned here, it is necessary first for the attacker to be able to determine how the data is chunked, i.e. to determine the points at which one chunk ends and the next chunk begins.  Depending on the software used and the capabilities of the attacker, this may be more or less difficult.

\subsection{Server access}

While encrypted backup software, by definition, encrypts the data being stored, in most (but perhaps not all) cases the server which is being used for storage is aware of the sizes of the chunks being stored.
This is useful for the efficient storage of deduplicated archives; if the last archive using a chunk of data is deleted, that data is no longer needed and can be removed from the backing storage providing that there is some way for the backing storage to identify the data in question (and distinguish it from other chunks uploaded during the same archive creation operation).

As such, the underlying storage system --- or anyone with access to it --- can in most cases easily determine the sequence of sizes of chunks constituting an archive.

\subsection{Fine-grained traffic analysis}
\label{subsection:traffic}

Even without access to the server, it may be possible to reveal the sequence of chunk sizes by monitoring network traffic.  Typically each chunk of data will be stored as a request to the server (potentially an HTTPS request over a new TCP connection for each chunk, but often one of a series of requests over a long-lived TCP connection).  As data is sent over the network, it will typically be sent as a series of maximum-segment-size\cite{RFC879} (MSS) TCP segments followed by a smaller segment with the "left over" data.  Even when a TCP connection is long-lived and used for many requests, if the network is fast enough to avoid having requests buffered on the client side, the individual requests can often be identified by the presence of a less-than-MSS TCP segment.

While this attack can obviously be carried out by an attacker who can read the (encrypted) network traffic between the client and server, the ability to read the traffic is not necessary (and indeed adds nothing assuming that the request transport is adequately secure, e.g. using TLS).  The requirement for carrying out this attack is merely the ability to measure {\em how much} traffic is being sent --- which for a system connected via a wireless network could be achieved by merely measuring the amount of radio traffic, even if the wireless network is cryptographically secured.

\subsection{Effects of compression}

Backup software typically compresses data being archived, in order to minimize storage usage, but for content-defined chunking to be used for deduplication purposes it's necessary that the chunking take place prior to data compression.  The previously-mentioned attacks reveal the {\em compressed} size of chunks (plus storage and network overheads), which may leave the actual uncompressed chunks ambiguous.

Data being archived typically consists of a mixture of incompressible data --- multimedia files and previously compressed data -- and data which compresses by anywhere between 2x and 10x.  In the former case, revealing the compressed size of a chunk will almost certainly reveal the length of the plaintext chunk; in the latter case, there may be several points where the chunk could have ended which yield the same compressed size, and thus several possibilities for the ``window'' which triggered the end of the chunk.

\subsection{Deduplication oracle}

Another attack is possible which can directly reveal the uncompressed chunk size by taking advantage of deduplication of archived data.  Recognizing that the decision of whether to end a chunk depends solely on data prior to the potential dividing point, if we know a chunk of data which has been previously stored, we can use the deduplication process as an oracle to answer the question ``does a chunk end at this point'' by causing the data {\tt {[query data][previously stored chunk]}} to be archived; a rough measurement of the total amount of new data stored then reveals whether {\tt {[previously stored chunk]}} remained as a chunk (and was deduplicated away), i.e. if the chunking process decided to end a chunk at the end of {\tt {[query data]}}.

Given this oracle and a known chunk {\tt X}, a chosen-plaintext attack can create archives {\tt aX}, {\tt abX}, {\tt abcX}, \ldots to determine the exact point where the data {\tt abc\ldots} will be chunked.  While this attack requires only coarse-grained traffic analysis --- or potentially none if the total volume of data stored is revealed in other ways (e.g. via billing records) --- and reveals the precise location where the plaintext is divided into chunks, it has the disadvantage compared to other attacks of being a chosen-plaintext attack and requiring a larger volume of data to be uploaded (potentially across many archives).

\end{document}